\documentclass[iop,numberedappendix,appendixfloats,twocolappendix]{emulateapj}

\usepackage{hyperref}
\usepackage{epsfig}
\usepackage{natbib}
\usepackage{graphicx}    
\usepackage{makeidx}
\makeindex                       
\usepackage{verbatim}                                   
\usepackage{amsmath,amsthm,amsfonts,amsopn,amssymb}     
\usepackage{longtable}
\usepackage{afterpage}
\usepackage{rotating} 
\usepackage{ulem}
\usepackage{epstopdf}  
\usepackage{multirow}
\usepackage{color}
\usepackage[left]{lineno}
\usepackage{amsmath}
\usepackage{bm}
\usepackage{afterpage}
\usepackage{chngpage}

\newcommand{\rprs}{R_{\rm{P}}/R_{*}}
\newcommand{\ars}{a/R_{*}}

\accepted{in AJ on March 15, 2016}

\slugcomment{}
\shorttitle{Planet Hunters K2 Follow-up}
\shortauthors{Schmitt et al. (2016)}

\begin{document}
\title{Planet Hunters. X. Searching for Nearby Neighbors of 75 Planet and Eclipsing Binary Candidates from the K2 \textit{Kepler} extended mission}
\author{
Joseph R. Schmitt\altaffilmark{1},
Andrei Tokovinin\altaffilmark{2},
Ji Wang\altaffilmark{3},
Debra A. Fischer\altaffilmark{1},
Martti H. Kristiansen\altaffilmark{4,5},
Daryll M. LaCourse\altaffilmark{5}, 
Robert Gagliano\altaffilmark{5}, 
Arvin Joseff V. Tan\altaffilmark{5}, 
Hans Martin Schwengeler\altaffilmark{5}, 
Mark R. Omohundro\altaffilmark{5}, 
Alexander Venner\altaffilmark{5}, 
Ivan Terentev\altaffilmark{5},  
Allan R. Schmitt\altaffilmark{5}, 
Thomas L. Jacobs\altaffilmark{5},
Troy Winarski\altaffilmark{5}, 
Johann Sejpka\altaffilmark{5}, 
Kian J. Jek\altaffilmark{5},
Tabetha S. Boyajian\altaffilmark{1},
John M. Brewer\altaffilmark{1},
Sascha T. Ishikawa\altaffilmark{6},
Chris Lintott\altaffilmark{7},
Stuart Lynn\altaffilmark{6,8},
Kevin Schawinski\altaffilmark{9}, 
Megan E. Schwamb\altaffilmark{10}, and
Alex Weiksnar\altaffilmark{6}
}
\email{joseph.schmitt@yale.edu}

\altaffiltext{1}{Department of Astronomy, Yale University, New Haven, CT 06511 USA}
\altaffiltext{2}{Cerro Tololo Inter-American Observatory, Casilla 603, La Serena, Chile}
\altaffiltext{3}{Department of Astronomy, California Institute of Technology, Pasadena, CA 91125, USA}
\altaffiltext{4}{DTU Space, National Space Institute, Technical University of Denmark, Elektrovej 327, DK-2800 Lyngby, Denmark}
\altaffiltext{5}{Planet Hunters citizen scientist}
\altaffiltext{6}{Adler Planetarium, Department of Citizen Science, 1300 S Lake Shore Dr, Chicago, IL 60605}
\altaffiltext{7}{Oxford Astrophysics, Denys Wilkinson Building, Keble Road, Oxford OX1 3RH}
\altaffiltext{8}{CartoDB, 247 Centre Street, New York, NY 10013, USA}
\altaffiltext{9}{Institute for Astronomy, Department of Physics, ETH Zurich, Wolfgang-Pauli-Strasse 27, CH-8093 Zurich, Switzerland}
\altaffiltext{10}{Institute for Astronomy and Astrophysics, Academia Sinica; 11F AS/NTU,  National Taiwan University, 1 Roosevelt Rd., Sec. 4, Taipei 10617, Taiwan}

\begin{abstract}

We present high-resolution observations of a sample of 75 K2 targets from Campaigns 1-3 using speckle interferometry on the Southern Astrophysical Research (SOAR) telescope and adaptive optics (AO) imaging at the Keck II telescope. The median SOAR $I$-band and Keck $K_s$-band detection limits at $1\arcsec$ were $\Delta m_{I}=4.4$~mag and $\Delta m_{K_s}=6.1$~mag, respectively.  This sample includes 37 stars likely to host planets, 32 targets likely to be eclipsing binaries (EBs), and 6 other targets previously labeled as likely planetary false positives.  We find nine likely physically bound companion stars within $3\arcsec$ of three candidate transiting exoplanet host stars and six likely EBs.  Six of the nine detected companions are new discoveries; one of the six, EPIC 206061524, is associated with a planet candidate. Among the EB candidates, companions were only found near the shortest period ones ($P<3$~days), which is in line with previous results showing high multiplicity near short-period binary stars.  This high-resolution data, including both the detected companions and the limits on potential unseen companions, will be useful in future planet vetting and stellar multiplicity rate studies for planets and binaries.

\end{abstract}


\section{Introduction}
\label{sec:intro}

The multiplicity of stellar systems has been well studied \citep{Duchene2013}, from M-dwarfs \citep[e.g.,][]{Fischer1992, Dieterich2012} to solar-type stars \citep[e.g.,][]{Abt1976, Tokovinin2014} to higher mass stars \citep[e.g.,][]{Garmany1980, Sana2012}.  High-resolution imaging is an effective method for searching for companion stars.  Adaptive optics (AO) is one such method, which uses natural or laser guiding stars to measure the air turbulence and deformable mirrors to correct for it, improving the angular resolution of astronomical images.  AO usually provides the highest resolution for ground-based methods outside of long baseline interferometry.  Speckle methods, on the other hand, take many images of the target star with millisecond exposures (a data cube), essentially freezing the air turbulence in place for the duration of the short observation, allowing for diffraction-limited resolution, as opposed to seeing-limited. With speckle interferometry, a Fourier analysis of every frame is performed to find nearby companions \citep[e.g.,][]{Howell2011}.  With lucky imaging, a subset of only the best frames are selected for analysis \citep[e.g,][]{Daemgen2009}. In this paper, we perform AO and speckle observations to search for companion stars to planet host stars or eclipsing binary (EB) candidates from the extended \textit{Kepler} mission (K2).  

\section{Target Selection}
\label{sec:targets}

The \textit{Kepler} mission \citep{Borucki2010} observed $\sim$160,000 stars almost continuously for nearly four years searching for planetary transits.  The mission discovered more than 1000 planets and another $\sim3700$ planet candidates\footnote{http://exoplanetarchive.ipac.caltech.edu, last accessed 2016 February 25} \citep{Coughlin2015}.  In 2013 May, the second of four reaction wheels on the \textit{Kepler} telescope failed, making it unable to continue observing the same field. In its two-wheel phase, called K2, the spacecraft can only reliably point at fields in the ecliptic plane for $\sim80$ day long campaigns before it must turn to a new field to avoid the Sun. The \textit{Kepler} spacecraft in its K2 mission continues to be a source of discovery for exoplanets \citep{Howell2014}.

Our target list consists of 75 stars observed by K2 during Campaigns 1-3. We conducted follow-up images of the 56 Campaign 1 (C1) targets and the two Campaign 2 (C2) targets at SOAR and observed the 17 Campaign 3 (C3) targets using Keck.  The targets and their designations are listed in Table~\ref{tab:bias}, which also lists the selection biases for each target.  The URLs within Table~\ref{tab:bias} contain the GO proposal identifier as well as the full proposal for each target.  The periods and epochs for all the EBs and EB candidates are listed in Table~\ref{tab:ebs}.  The planet candidates are discussed further in Section~\ref{sec:tap}. 

\begin{deluxetable*}{ccccccccc}
\tablewidth{0pt}
\tablecaption{EB and EBC properties.}
\tablehead{
\colhead{EPIC}                     & 
\colhead{Campaign}                 & 
\colhead{Status}                   &  
\colhead{$P_{\rm{PH}}$}            & 
\colhead{Epoch$_{\rm{PH}}$}        &  
\colhead{$P_{\rm{KEB}}$}           & 
\colhead{Epoch$_{\rm{KEB}}$}       \\ 
\colhead{ID}                       & 
\colhead{}                         &
\colhead{}                         &  
\colhead{(days)}                   & 
\colhead{(KBJD\tablenotemark{a})}  & 
\colhead{(days)}                   & 
\colhead{(KBJD)}                   }
\startdata
201160662\phn               & C1  & EBC        & \phn1.537          & 1975.957  & $\phn1.53687\pm0.00013$              & 1981.05912\phd$\pm$\phd\nodata\phn  \\ 
201207683\phn               & C1  & EBC        & \tablenotemark{b}  & 2002.312  & \phd\nodata                          & \phn\phn\phd\nodata                 \\ 
201246763\phn               & C1  & EBC        & 43.663             & 2014.326  & 43.66300\phd$\pm$\phd\nodata\phn     & $1962.93272\pm0.57131$              \\ 
201253025\tablenotemark{c}  & C1  & EB\phm{C}  & \phn3.392          & 1980.767  & $\phn6.78617\pm0.00105$              & $1978.12149\pm0.03787$              \\ 
201270464\tablenotemark{d}  & C1  & EBC        & \phn3.155          & 1977.436  & \phd\nodata                          & \phn\phn\phd\nodata                 \\ 
201324549\tablenotemark{e}  & C1  & EBC        & \phn2.519          & 1979.500  & \phd\nodata                          & \phn\phn\phd\nodata                 \\ 
201407812\phn               & C1  & EBC        & \phn2.827          & 1979.490  & $\phn2.82678\pm0.00030$              & $1984.22530\pm0.04365$              \\ 
201408204\phn               & C1  & EB\phm{C}  & \phn8.482          & 2024.606  & $\phn8.48191\pm0.00137$              & $2025.34343\pm0.03497$              \\ 
201458798\phn               & C1  & EBC        & \phn0.619          & 1977.568  & $\phn0.61939\pm0.00003$              & \phn\phn\phd\nodata                 \\ 
201488365\phn               & C1  & EB\phm{C}  & \phn3.362          & 1975.859  & $\phn3.36426\pm0.00039$              & $1981.44082\pm0.04704$              \\ 
201567796\phn               & C1  & EBC        & \phn5.011          & 1979.536  & $\phn5.00861\pm0.00069$              & $2003.31875\pm0.03142$              \\ 
201576812\phn               & C1  & EB\phm{C}  & \phn5.730          & 1975.858  & $\phn5.72823\pm0.00084$              & $1989.66917\pm0.02229$              \\ 
201594823\phn               & C1  & EB\phm{C}  & \phn1.301          & 1976.659  & $\phn1.30062\pm0.00010$              & $1977.93351\pm0.01931$              \\ 
201626686\phn               & C1  & EBC        & \phn5.280          & 1979.356  & $\phn5.28011\pm0.00069$              & $1973.08643\pm0.03258$              \\ 
201648133\phn               & C1  & EBC        & 35.020             & 1980.807  & $35.02000   \pm0.00735$              & 1972.47647\phd$\pm$\phd\nodata\phn  \\ 
201665500\phn               & C1  & EB\phm{C}  & \phn3.054          & 1977.539  & $\phn3.05352\pm0.00033$              & $1990.67896\pm0.03027$              \\ 
201704541\phn               & C1  & EB\phm{C}  & \phn0.411          & 1976.547  & $\phn0.41138\pm0.00002$              & $1975.24518\pm0.02477$              \\ 
201705526\phn               & C1  & EBC        & 18.103             & 1986.610  & $18.09409   \pm0.00381$              & 2012.62636\phd$\pm$\phd\nodata\phn  \\ 
201711881\phn               & C1  & EB\phm{C}  & \phn5.468          & 1977.988  & $\phn5.46846\pm0.00077$              & $1975.43501\pm0.30923$              \\ 
201725399\phn               & C1  & EBC        & \phn2.162          & 1978.253  & $\phn2.16127\pm0.00020$              & $1986.34269\pm0.04807$              \\ 
201826968\phn               & C1  & EBC        & \phn0.367          & 1976.608  & $\phn0.36176\pm0.00002$              & $1974.23489\pm0.03098$              \\ 
201890494\phn               & C1  & EBC        & \phn2.536          & 1977.446  & $\phn2.53657\pm0.00026$              & $1964.73129\pm0.02028$              \\ 
201928968\phn               & C1  & EBC        & \phn0.320          & 1980.390  & $\phn0.32000\pm0.00001$              & 1979.66097\phd$\pm$\phd\nodata\phn  \\ 
203533312\tablenotemark{f}  & C2  & EBC        & \phn0.176          & 2061.640  & \phd\nodata                          & \phn\phn\phd\nodata                 \\ 
204129699\phn               & C2  & EBC        & \phn1.258          & 2060.600  & $\phn1.25780\pm0.00010$              & 2061.86700\phd$\pm$\phd\nodata\phn  \\ 
205985357\phn               & C3  & EBC        & \phn4.128          & 2148.728  & \phd\nodata                          & \phn\phn\phd\nodata                 \\ 
206029314\phn               & C3  & EBC        & \phn7.026          & 2148.069  & \phd\nodata                          & \phn\phn\phd\nodata                 \\ 
206047297\phn               & C3  & EBC        & 27.317	            & 2166.457  & \phd\nodata                          & \phn\phn\phd\nodata                 \\ 
206135075\phn               & C3  & EBC        & 54.976	            & 2149.868  & \phd\nodata                          & \phn\phn\phd\nodata                 \\ 
206135267\phn               & C3  & EB\phm{C}  & \phn2.533          & 2147.052  & \phd\nodata                          & \phn\phn\phd\nodata                 \\ 
206152015\phn               & C3  & EBC        & \phn0.809          & 2147.088  & \phd\nodata                          & \phn\phn\phd\nodata                 \\ 
206173295\phn               & C3  & EBC        & \phn2.176          & 2147.784  & \phd\nodata                          & \phn\phn\phd\nodata                 \\ 
206311743\phn               & C3  & EBC        & \phn4.312        & 2155.042  & \phd\nodata                          & \phn\phn\phn\nodata                 \\
206380678\tablenotemark{e}  & C3  & EBC        & \phn2.271          & 2147.270  & \phd\nodata                          & \phn\phn\phd\nodata
\enddata
\tablecomments{We list here the periods and epochs of the EBs and EB candidates both estimated by PH users, $P_{\rm{PH}}$ and Epoch$_{\rm{PH}}$, and many of them also with data from a preliminary \textit{Kepler} Eclipsing Binary catalog \citep[][K. Conroy 2015, private communication]{Prsa2011}, $P_{\rm{KEB}}$ and Epoch$_{\rm{KEB}}$.}
\tablenotetext{a}{\textit{Kepler} Barycentric Julian Day (KBJD) is equal to JD minus 2454833.0 (UTC$=$2009 January 1 12:00:00).}
\tablenotetext{b}{Single stellar eclipse (depth~$\sim24\%$).}
\tablenotetext{c}{PH users counted each transit as a primary transit, while the initial KEB catalog counted primary and secondary events. This explains the factor of two difference in the periods and the offset in the epoch.}
\tablenotetext{d}{Eclipse profile is shallow and V-shaped. May have alternating minima.}
\tablenotetext{e}{Eclipse profile is shallow and V-shaped.}
\tablenotetext{f}{The EB candidate may have $P_{\rm{PH}}=0.361$ days, double what is listed in the table.}
\label{tab:ebs}
\end{deluxetable*}

\subsection{Planet Hunters Targets}

The citizen science project Planet Hunters\footnote{\url{http://www.planethunters.org/}} \citep[PH,][]{Fischer2012} was the primary source for finding 45 targets from C1 to C3.  PH is a member of the citizen science Zooniverse\footnote{\url{https://www.zooniverse.org/}} project \citep{Lintott2008}.  PH volunteers organized their search on their own, surveying data from the K2 self-flat fielding (K2SFF) database \citep{Vanderburg2014} or reducing the data themselves with the Guest Observer software PyKE \citep{Still2012} or their own, self-created tools (e.g., LcTools{\small \textcopyright}\footnote{\url{https://dl.dropboxusercontent.com/u/78120543/LcTools/LcTools\%20Product\%20Description.htm}}).  Users check light curves for the signature of a planetary transit, EB, or other astrophysical objects \citep[e.g.,][]{Kato2014}.  This project crowd-sources the analysis of K2 light curves and has been successful in the past in finding planet candidates \citep{Fischer2012, Lintott2013, Wang2013, Schmitt2014a, Wang2015b}, confirming planets \citep{Schwamb2013, Wang2013, Schmitt2014b}, finding EBs \citep{LaCourse2015}, and finding other, as of yet, unidentified signals \citep{Boyajian2016}.

Among these 45 PH targets are WASP-85A~b \citep{Brown2015}, which is a known exoplanet in a binary system, and nine other targets known to be EBs (eight from C1 and one from C3), according to the Guest Observer (GO) proposals requesting the targets.  We have classified 10 of the 45 PH targets as Planet Hunter Objects of Interest (PHOIs), which is analagous to the Kepler Objects of Interest (KOI) designation.  These were also discovered independently by \citet{Vanderburg2016}.  The rest of the PH targets are either previously known EBs or newly discovered candidate EBs, of which many were also independently found in \citet{Armstrong2016}.

\subsection{\citet{Foreman-Mackey2015} Targets}

Of the C1 targets, 30 were selected from the K2 C1 planet candidate list by \citet{Foreman-Mackey2015}, which comprises 36 planet candidates orbiting 31 stars.  Several of these were also noted by PH volunteers.  We selected all but one star, excluding EPIC 201565013 owing to its faintness, $K_P=16.91$~mag.  Of the 30 stars obtained from \citet{Foreman-Mackey2015}, one of the targets, EPIC 201505350 (K2-19), was later confirmed to host a planet using ground-based photometric follow-up, transit timing analysis, AO imaging, spectroscopy, and photo-dynamical analysis \citep{Armstrong2015,Barros2015,Narita2015}. \citet{Montet2015} later validated planets around 16 of these 30 stars, including the previously mentioned K2-19, using a statistical elimination of astrophysical false positives, while deeming six others to be likely false positives. EPIC 201465501 (K2-9) was also independently validated by \citet{Schlieder2016}.  We observed all 30 of these targets, regardless of their designation.  

\section{Observations and Data Reduction}
\label{sec:obs}

On the nights of 2015 May 2-3, we observed 58 stars from the K2 program; 56 were from C1 and two were from C2.  We used speckle interferometry with HRCAM \citep{Tokovinin2008}, a high-resolution camera on the SOAR Adaptive Optics Module (SAM) at the 4.1-meter Southern Astrophysical Research (SOAR) telescope at Cerro Pach\'on Observatory.  On the night of 2015 July 29, a portion of the night was devoted to observing the 17 stars from C3 with the NIRC2 instrument on the Keck II telescope.  

\subsection{SOAR speckle interferometry}

For the 58 targets observed by SOAR, we used the Bessel $I$-band filter (central wavelength = 866.5 nm) on HRCAM because this provided better seeing and a wider bandwidth (FWHM$=391.4$~nm) than in the visual and favored the detection of M-dwarf companion stars.  Some time was lost because of clouds and technical problems.  For both nights, the telescope was pointed directly into a strong wind.  This buffeted the telescope and could cause high jitter up to $3\arcsec$.

For each target star, we typically took four data cubes with 400 images each.   For the first two cubes, the field size was $6\farcs092\times6\farcs092$ using 200x200 binned (2x2) pixels with typical exposures of 200 ms.  In the last two cubes, we did not bin the data.  Correspondingly, the field size shrank to $3\arcsec\times3\arcsec$.  The exposure times ranged between 20 and 50 ms for the smaller field.  For the highest wind conditions, we only collected binned pixel data cubes. The wider fields allowed for the detection of fainter, more distant companions, while the narrow field cubes allow for the detection of brighter, closer companions.  The detected companions have all been confirmed in multiple data cubes.

\begin{figure*}[htb]
\includegraphics[width=1.00\textwidth]{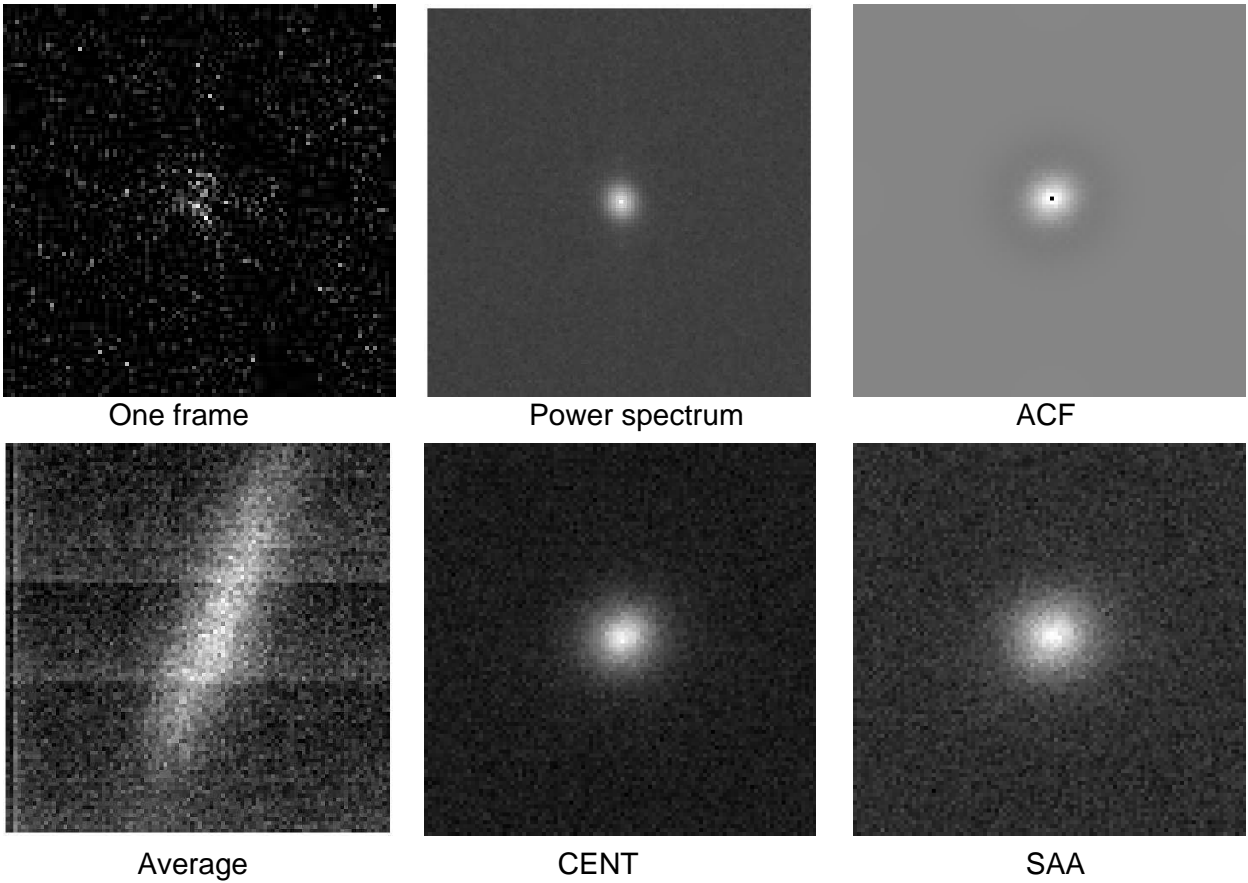}
\caption{Various data products produced by the reduction pipeline for a single star (EPIC 201555883), i.e., a non-detection.  The top-left image is what a single frame in the analysis looks like, while the other five images show composites of all frames:  the power spectrum, auto-correlation function (ACF), the average image, the average image re-centered around the centroid (CENT), and the shift-and-add method (SAA) of centering the image on the brightest pixel.  The large jitter in the average image is caused by wind buffeting the telescope.  The scale of each image is $6\farcs092\times6\farcs092$.}
\label{fig:images}
\end{figure*}

The data were processed using a standard speckle pipeline \citep{Tokovinin2010}.  The pipeline delivered five products for each target: a power spectrum, an auto-correlation function computed from the power spectrum, an average image, an average image re-centered around the centroid, and a shift-and-add image re-centered on the brightest pixel.  See Figure~\ref{fig:images} for an example of each image product for EPIC 201555883.  The detector orientation and pixel scale were accurately calibrated on wide binaries with well-modeled linear motions.

The faint magnitudes of our target stars required modifications to the standard pipeline.  Hot pixels from longer than standard exposures were fixed by removing the dark current and the bias and accounting for the 2x2 binning.  Clock-injection charges (CICs) were a major contributor to the power spectrum of faint stars.  CICs create a background of spurious photon spikes that bias the centroid of the star toward the middle of the frame, a problem for frames in which the star deviated far from the center, which occurred during periods of high wind.  The CICs were removed by smoothing the images with a width of five pixels, taking its median-average as the background and then subtracting it.  A threshold of 0.3 times the maximum intensity above the background was also subtracted (and clipped at zero).  This produces a properly centered image.  To reduce the noise associated with CICs, we multiplied each re-centered image by a Gaussian mask of 15 pixel ($0\farcs46$) FWHM and calculated alternative power spectra from those masked images. Masking improved the signal-to-noise ratio in the power spectrum, making closer companions more detectable at the expense of reducing detectability of companions beyond $0\farcs5$.  Since high wind resulted in temporary losses of the image from the field of view (FOV), we removed frames if the centroid was calculated to be within 20 pixels of the frame border. For the shift-and-add method, if the brightest pixel was more than 20 pixels away from the centroid, the frame was removed as a likely cosmic ray event. 

\subsection{Keck AO imaging}

We observed 17 K2 C3 planet candidates with the NIRC2 instrument at the Keck II telescope (Mauna Kea, Hawaii, United States). NIRC2 is a near infrared imager designed for the Keck AO system \citep{Wizinowich2000}. The observations were made on UT 2015 July 29, with $0\farcs8$-$1\farcs0$ seeing. We selected the narrow camera mode, which has a pixel scale of 10 mas/pixel. The FOV is thus $10\arcsec\times10\arcsec$ for a mosaic 1K$\times$1K detector. All images were taken in the $K_s$ band, which provides higher sensitivity than $J$ and $H$ band for bound companions with late spectral type.  Among the many sensors that allow the primary mirror segments to act as one mirror, an error in one of the sensors caused a co-phasing issue with about $25\%$ of the mirror segments. The Keck team hopes to implement better alarms on the primary mirror to alert them to similar mirror-induced image quality problems (J. Lyke 2016, private communication).  This degraded our AO-corrected point spread function and decreased our performance relative to standard NIRC2 observations.  Exposure time was set such that the peak flux of the target was at least 5000 ADU after co-add. Before co-add, peak flux was limited to 2000 ADU to avoid nonlinearity and detector persistence.  We used a 3-point dither pattern (three corners of a square) with a throw of $2\farcs5$.  We avoided the lower left quadrant in the dither pattern because it has a much higher instrumental noise than the other three quadrants on the detector.

The raw Keck NIRC2 AO data were processed using standard techniques to replace bad pixels, subtract dark frames, flat-field, subtract sky background, and align and co-add frames. Our own custom program recorded the differential magnitude, separation, position angle ($\theta$), and detection significance.  All detections were then visually checked to remove confusions such as speckles, background extended sources, and cosmic ray hits. 

\section{Transit fitting for PHOIs}
\label{sec:tap}

\begin{figure*}
\includegraphics{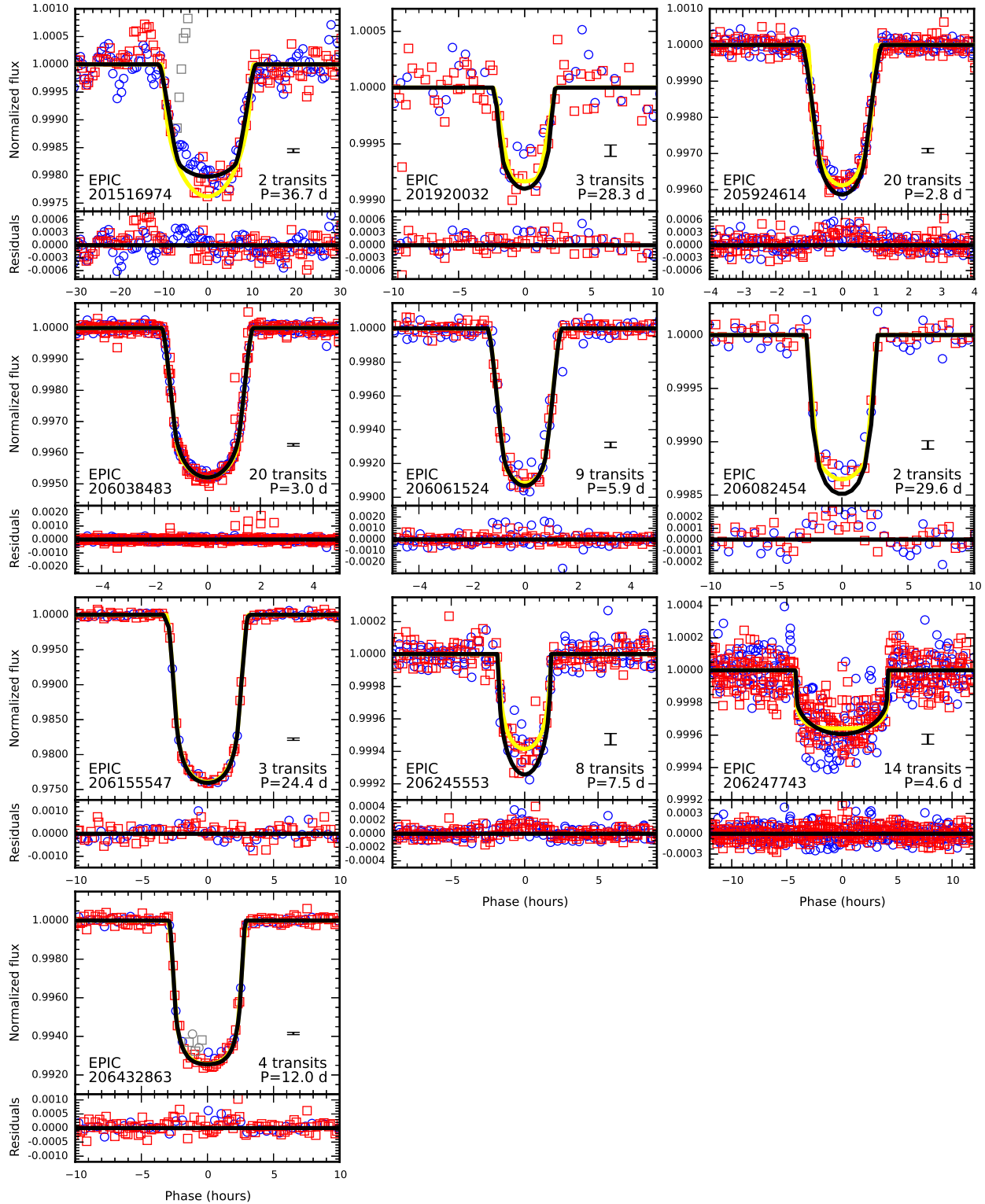}
\caption{Top panels are the transit fits for all PHOIs.  Odd-numbered transits (one-based indexing) are shown by red squares, while even-numbered transits are shown by blue circles.  The solid, black line is the median model fit, while the most likely individual model from the MCMC analysis is highlighted in yellow.  Visible differences between these two fit lines are caused by the bimodality of the MCMC results.  A representative error bar ($\pm\sigma$) is shown in black in the bottom right of each panel above the label for number of transits. Grayed squares for EPIC 201516974 represent data points masked due to detrending issues, while grayed circles and squares for EPIC 206432863 represent data points masked due to suspected starspot crossings.  The bottom panels display the residuals of the data minus the median model.  Comparing to the median model results in some structure in the residuals when there is a significant difference between the median model and the most likely model.}
\label{fig:transits}
\end{figure*}

For each of the PHOIs listed in Table~\ref{tab:tap}, we downloaded the K2SFF light curves from \citet{Vanderburg2014}.  The K2SFF data reduction process removes the effect of the spacecraft thruster fires that occur approximately every six hours, although it does not do so perfectly in all cases.  The K2SFF reduction process is not intended to remove stellar variations.  We flattened these K2SFF light curves with low-order ($n\leq 4$) polynomial fits to out-of-transit data and clipped the light curves around the transits using a combination of our own codes, the aforementioned PyKE software, and the IDL program \texttt{autoKep} \citep{Gazak2012}.  One occasional side effect of the K2SFF reduction process was a ringing-like signature in the location where a transit should have been. These affected transits were typically removed from our analysis.  However, in the three-transit case of EPIC 206155547, we extracted one of these badly reduced transits from the raw data since there was no apparent data discontinuity caused by a thruster fire during the transit.  We then fit the raw, out-of-transit data on either side of the transit to a quadratic polynomial and removed the trend in a similar manner as the flattening of the K2SFF light curves.  Another common effect was a spike in brightness within transits, which typically degraded the transit to such a degree that the transit was simply removed from the analysis.  One exception is the brightness spike in the first transit of EPIC 201516974.  Due to its longer period ($P=36.7$~day) and thus longer duration, the spike degraded only a minority of the transit. Therefore, we simply masked the spike out (partially shown by gray squares in Figure~\ref{fig:transits}).

Some of the PHOIs have suspected signals of stellar activity, either from large-scale brightness variations in the overall light curve or from bumps within the transit. For EPIC 206432863, we masked out two suspected starspot crossings (shown in gray in Figure~\ref{fig:transits}).  For other stars, there were no sharp, clearly defined starspot or plage crossings, but small-scale stellar activity was evident in the increased scatter in the in-transit residuals of some of the fits.  

The transit parameters were fit by the IDL program \texttt{TAP} \citep{Gazak2012}, an MCMC fitting routine using EXOFAST \citep{Eastman2013} to calculate \citet{Mandel2002} transit models using a wavelet-based likelihood function \citep{Carter2009}.  \texttt{TAP} was used to fit the impact parameter $b$, the transit duration $T$, the ratio of the planet radius to the stellar radius $\rprs$, the midtransit times, linear and quadratic limb darkening, red and white noise, and the coefficients of a quadratic normalization polynomial for each individual transit event (in case of an imperfectly normalized or flattened light curve).  The ratio of the semimajor axis to the stellar radius $\ars$ and the inclination $i$ were derived from the posterior of each solution by \texttt{TAP} using $T$ and $b$.  Circular orbits were assumed.  Each set of transits were fit with ten MCMC chains of various lengths (100,000-2,000,000) to ensure no indication of non-convergence according to the Gelman-Rubin statistic \citep{Gelman1992}.  The period $P$ is poorly constrained by \texttt{TAP}. Therefore, for each PHOI, we randomly drew 1,000,000 samples of each individual transit's midpoint from the \texttt{TAP} posterior and calculated the period between consecutive fitted transits, taking into account missing transits where necessary.  We then took the median and its $1\sigma$ upper and lower error bars.  The transits and their fits are shown in Figures~\ref{fig:transits}.

One important caveat to the numbers in Table~\ref{tab:tap} and the best-fit lines in Figure~\ref{fig:transits} is that we have chosen to present the median values and their $1\sigma$ error bars because they better capture the distribution of each parameter.  However, the median value is not necessarily the most likely model.  Transit light curve fitting can result in bimodal distributions due to weak degeneracies between the parameters, such as $T$ and $b$.  More often than not, the effect is minor, and the median value closely approximates the most likely value for the most important physical parameters, such as $\rprs$.  However, there are cases where the most likely value is moderately different from the median, even being at the edge of the $1\sigma$ error bars in the more extreme cases.  In Figure~\ref{fig:transits}, this causes the structure one sees in the residual to the median model.  One specific example of this effect is the fit for EPIC 206082454 (PHOI-6~b).  The median value of $\rprs$ is actually a local minimum.  For this planet, the upper and lower $1\sigma$ limit closely approximates the center of the two local maxima.  This has the effect in Figure~\ref{fig:transits} of placing the fit line below most of the data points in the bottom of the transit.  The same applies for EPIC 206245553 (PHOI-8~b). For both stars, $\ars$ and $i$ also show a bimodal distribution.  To qualitatively show the agreement (or disagreement) between the median model and the single most likely  individual model, we also plot the most likely single model from the MCMC analysis in yellow.  

\begin{deluxetable*}{lcccccccc}
\tablewidth{0pt}
\tablecaption{PHOI transit fit results and derived parameters.}
\tablehead{
\colhead{EPIC}                     & 
\colhead{PHOI}                     & 
\colhead{$P$}                      & 
\colhead{Epoch}                    &  
\colhead{$T$}                      & 
\colhead{$b$}                      & 
\colhead{$\rprs$}                  & 
\colhead{$\ars$}                   & 
\colhead{$i$}                      \\
\colhead{ID}                       & 
\colhead{\phm{\tablenotemark{a}}ID\tablenotemark{a}} & 
\colhead{(days)}                   & 
\colhead{(KBJD\tablenotemark{b})}  & 
\colhead{(hr)}                  & 
\colhead{}                         & 
\colhead{}                         & 
\colhead{}                         & 
\colhead{(degrees)}                 }
\startdata
201516974 & \phn1 b & $36.7099_{-0.0126}^{+0.0124}$ & $1986.8056_{-0.0095}^{+0.0098}$ & $0.736_{-0.039}^{+0.070}$ & $0.90_{-0.12}^{+0.03}$ & $0.0489_{-0.0033}^{+0.0028}$ & $\phn6.9_{-\phn0.7}^{+\phn2.2}$ & $82.6_{-1.1}^{+2.5}$ \\
201920032 & \phn2 b & $28.2717_{-0.0139}^{+0.0141}$ & $2000.2058_{-0.0059}^{+0.0051}$ & $0.171_{-0.010}^{+0.011}$ & $0.01_{-0.75}^{+0.74}$ & $0.0264_{-0.0020}^{+0.0047}$ & $42.8_{-18.3}^{+\phn8.2}$ & $89.3_{-1.3}^{+0.5}$ \\
205924614 & \phn3 b & $\phn2.8493_{-0.0015}^{+0.0013}$ & $2150.4245_{-0.0008}^{+0.0008}$ & $0.078_{-0.002}^{+0.001}$ & $0.14_{-0.61}^{+0.49}$ & $0.0574_{-0.0019}^{+0.0032}$ & $10.1_{-\phn1.6}^{+\phn1.8}$ & $87.5_{-2.0}^{+1.7}$ \\
206038483 & \phn4 b & $\phn3.0026_{-0.0012}^{+0.0012}$ & $2149.0598_{-0.0005}^{+0.0005}$ & $0.120_{-0.001}^{+0.001}$ & $0.80_{-0.02}^{+0.01}$ & $0.0696_{-0.0008}^{+0.0010}$ & $\phn4.9_{-\phn0.5}^{+\phn0.5}$ & $80.6_{-1.2}^{+0.9}$ \\
206061524 & \phn5 b & $\phn5.8797_{-0.0015}^{+0.0018}$ & $2153.3239_{-0.0006}^{+0.0006}$ & $0.093_{-0.003}^{+0.003}$ & $0.80_{-0.03}^{+0.03}$ & $0.0982_{-0.0020}^{+0.0024}$ & $12.1_{-\phn0.8}^{+\phn0.9}$ & $86.2_{-0.4}^{+0.4}$ \\
206082454 & \phn6 b & $29.6260_{-0.0017}^{+0.0016}$ & $2160.5402_{-0.0011}^{+0.0011}$ & $0.194_{-0.004}^{+0.006}$ & $0.09_{-0.86}^{+0.74}$ & $0.0348_{-0.0022}^{+0.0036}$ & $36.8_{-11.9}^{+\phn9.9}$ & $89.0_{-1.0}^{+0.7}$ \\
206155547 & \phn7 b & $24.3872_{-0.0012}^{+0.0010}$ & $2152.8841_{-0.0002}^{+0.0002}$ & $0.226_{-0.002}^{+0.001}$ & $0.29_{-0.63}^{+0.07}$ & $0.1401_{-0.0013}^{+0.0014}$ & $32.4_{-\phn0.6}^{+\phn0.6}$ & $89.4_{-0.1}^{+0.1}$ \\
206245553 & 8 b & $\phn7.4950_{-0.0069}^{+0.0084}$ & $2154.6728_{-0.0018}^{+0.0013}$ & $0.147_{-0.005}^{+0.003}$ & $0.28_{-1.01}^{+0.57}$ & $0.0239_{-0.0021}^{+0.0035}$ & $11.8_{-\phn3.7}^{+\phn3.6}$ & $86.7_{-2.9}^{+2.4}$ \\
206247743 & 9 b & $\phn4.6028_{-0.0289}^{+0.0342}$ & $2147.8210_{-0.0041}^{+0.0047}$ & $0.341_{-0.009}^{+0.008}$ & $0.04_{-0.55}^{+0.62}$ & $0.0178_{-0.0006}^{+0.0012}$ & $\phn3.8_{-\phn0.7}^{+\phn0.5}$ & $83.4_{-6.9}^{+4.6}$ \\
206432863 & 10 b & $11.9897_{-0.0012}^{+0.0008}$ & $2150.8263_{-0.0005}^{+0.0005}$ & $0.223_{-0.002}^{+0.001}$ & $0.29_{-0.78}^{+0.22}$ & $0.0787_{-0.0011}^{+0.0010}$ & $15.1_{-\phn0.7}^{+\phn0.8}$ & $88.2_{-0.3}^{+0.4}$ 
\enddata
\tablecomments{For definitions of each parameter, see Section~\ref{sec:tap}.  These are the median and $1\sigma$ values from the \texttt{TAP} fits, which are not necessarily the best-fit models.  See Section~\ref{sec:tap} for a further discussion on using the median vs. the best-fit.}
\tablenotetext{a}{Full PC names are PHOI-1~b, PHOI-2~b, etc.}
\tablenotetext{b}{\textit{Kepler} Barycentric Julian Day (KBJD) is equal to JD minus 2454833.0 (UTC$=$2009 January 1 12:00:00).}
\label{tab:tap}
\end{deluxetable*}

\section{Results}
\label{sec:results}

\subsection{Previously known binaries}

For all 75 targets, we searched the literature for companions within our FOV.  Our search included several surveys and catalogs:  APASS \citep{Henden2014}, SDSS \citep{Alam2015}, 2MASS \citep{Strutskie2006}, \textit{WISE} \citep{Wright2010}, and the Washington Double Star (WDS) catalog \citep{Mason2001}.  Many of the potential companions were low signal to noise, had aberrations caused by diffraction spikes (particularly in SDSS), or were otherwise unlikely to be true stars.  We performed a manual triage to include only high-quality detections of companion stars. Unfortunately, however, the two C2 targets and 11 of the 17 C3 targets have not been observed by SDSS.  We identified four known companions in the literature search, one in SDSS (EPIC 201890494), one in the WDS catalog (EPIC 201862715), and two in \citet[][EPIC 201546283 and EPIC 201828749]{Montet2015}. 

The companion to EPIC 201890494 found by SDSS was successfully recovered.  We also recovered the companion to EPIC 201862715 (WASP-85).  This is a visual, G-K dwarf binary system \citep{Burnham1882} listed in the WDS catalog.  The primary component hosts an inflated hot Jupiter, named WASP-85A~b, which was confirmed via ground-based photometry, radial velocities, and K2 photometry \citep{Brown2015}. 

The two other stars known to have companions were discovered by \citet{Montet2015}, who observed seven of the candidates in \citet{Foreman-Mackey2015} with the Palomar High Angular Resolution Observer (PHARO) infrared detector \citep{Hayward2001} AO system \citep{Dekany2013} at the 5.1 meter Palomar Hale telescope.  The two of their targets that resulted in a detection of a nearby companion star were EPIC 201546283 and EPIC 201828749.  We recovered only the latter companion, which was originally measured to have $\rho=2\farcs46\pm0\farcs04$ and $\Delta m_J=1.462\pm0.012$~mag (B. Montet 2015, private communication).  The unrecovered companion was likely missed due to the combination of the companion's distance from the primary ($\rho=2\farcs98\pm0\farcs05$), putting it at the edge of our detector, and its faintness ($\Delta m_{K_s}=3.72$~mag, B. Montet 2015, private communication), which implies a higher $\Delta m_I$ close to our detection limit of $\Delta m_I=5.0$ at $>2\arcsec$ for this star.  

\begin{figure*}[htb]
\includegraphics[width=1.00\textwidth]{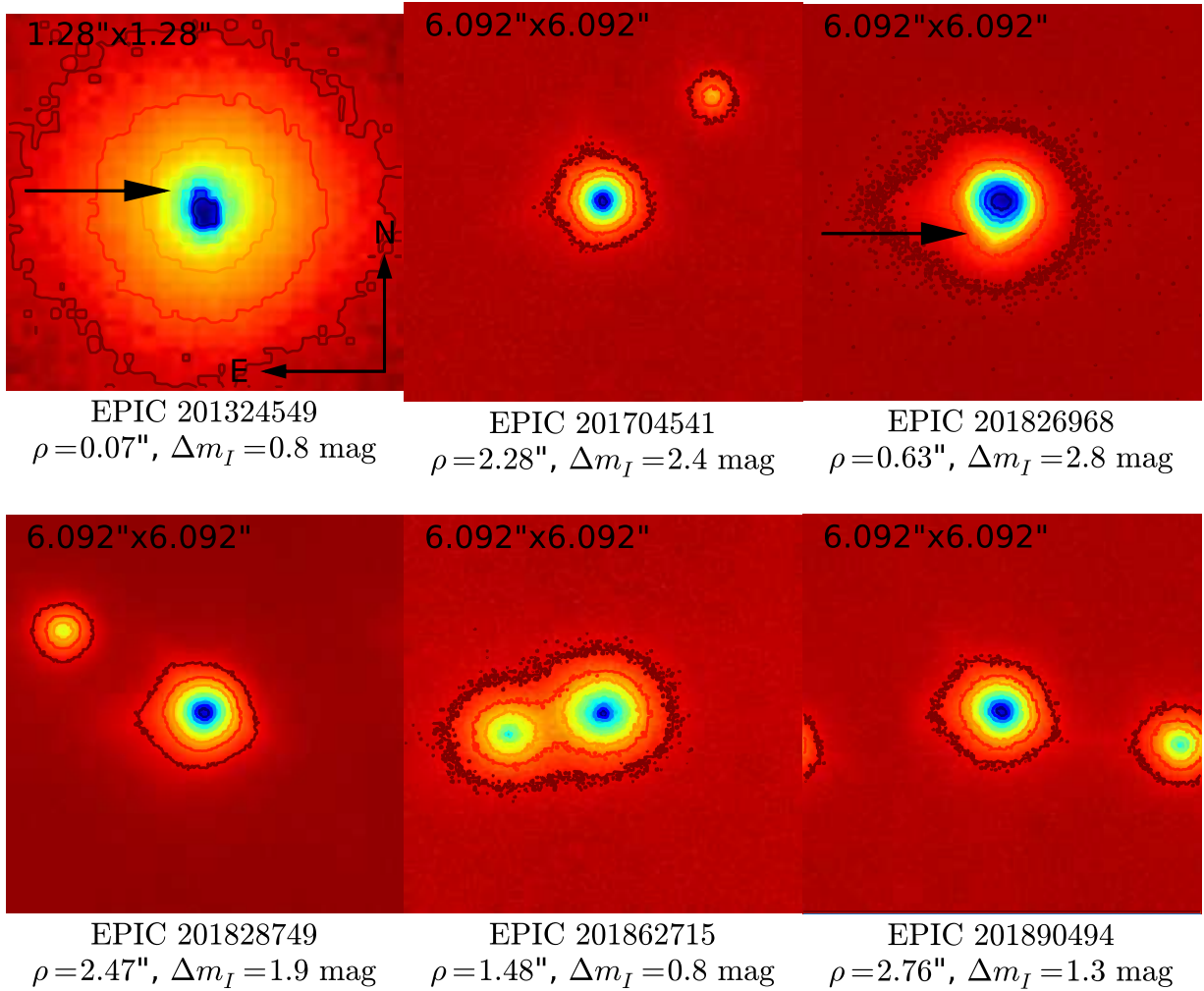}
\caption{Binaries detected by SOAR speckle interferometry.  Arrows point to the sub-arcsecond detections. The companion to EPIC 201324549 can be seen in the blue bump to the north-northeast, while the companion to EPIC 201826968 can be seen as the yellow bump south of EPIC 201826968.  These images are for illustrative value only.  They were not used to make the discovery. Each companion was independently detected in multiple data cubes.}
\label{fig:SOARdetections}
\end{figure*}

All three companions recovered were found with the SOAR observations.  See Table~\ref{tab:detect} for their properties and Figure~\ref{fig:SOARdetections} for images of the companions.  These images are for illustrative value only and were not used to make the discovery.  Each companion was found independently in multiple data cubes.

\subsection{New Detections}
\label{sec:Detect}

We detected three new companions with the SOAR observations (see Table~\ref{tab:detect}).  From the SOAR observations, we did not discover any new stellar companions among the planet candidates. Of the eight known EBs, one new companion was discovered (EPIC 201704541).  Around the 17 EB candidates, two new companion stars were discovered near EPIC 201324549 and EPIC 201826968.  See Figure~\ref{fig:SOARdetections} for speckle images of the companions.  Again, these images are for illustrative value only.  Each new companion was found in multiple data cubes.

\begin{figure*}[htb]
\includegraphics[width=1.00\textwidth]{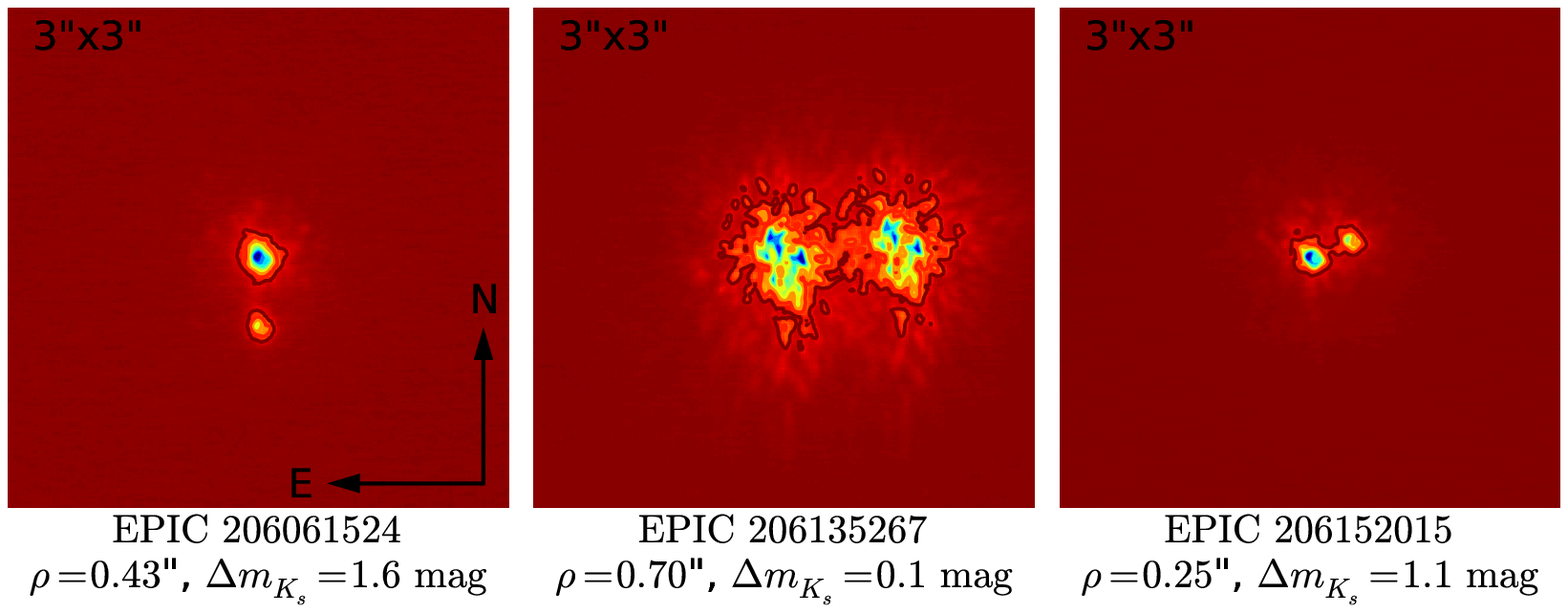}
\caption{The binaries detected by Keck NIRC2 AO imaging.  Due the to close separation of the two stars for EPIC 206135267 and their near equal brightness, the AO corrections system became confused, and the AO correction has been much reduced.}
\label{fig:keckdetections}
\end{figure*}

We also detected three new companions with the Keck observations (again, Table~\ref{tab:detect}).  From the Keck AO imaging, all three newly discovered companions are within $1\arcsec$ of the primary star. One companion was near one of the PHOIs (EPIC 206061524), one was near the EB (EPIC 206135267), and one was near an EB candidate (EPIC 206152015).  See Figure~\ref{fig:keckdetections} for the AO detection images. 

\begin{deluxetable*}{clcccccccccc}
\tablewidth{0pt}
\tablecaption{Binary detections.}
\tablehead{
\colhead{EPIC}                                    & 
\colhead{Status}                                  &
\colhead{First}                                   & 
\colhead{Epoch}                                   &   
\colhead{$\theta$}                                & 
\colhead{$\rho\sigma_{\theta}$\tablenotemark{a}}  & 
\colhead{$\rho$}                                  & 
\colhead{$\sigma\rho$}                            & 
\colhead{$\Delta m$}\tablenotemark{b}             & 
\colhead{Prob. Proj.\tablenotemark{c}}            \\
\colhead{ID}                                      & 
\colhead{}                                        &
\colhead{Detection}                               &    
\colhead{$+2000$}                                 & 
\colhead{(degrees)}                               & 
\colhead{(mas)}                                   & 
\colhead{($\arcsec$)}                             & 
\colhead{(mas)}                                   & 
\colhead{(mag)}                                   &
\colhead{(\%)}                                    }
\startdata
201324549   & EBC   & This paper (SOAR speckle)  & 15.3380  & \phn12.2   & \phn3.9  & 0.0721        & \phn4.4     & 0.8  & 0.000039              \\
201704541   & EB    & This paper (SOAR speckle)  & 15.3380  & 310.8      & \phn4.9  & 2.2793        & \phn4.9     & 2.4  & 0.041\phn\phn \phn    \\
201826968   & EBC   & This paper (SOAR speckle)  & 15.3379  & 164.7      & 13.3     & 0.6330        & \phn6.7     & 2.8  & 0.041\phn\phn\phn     \\
201828749   & PC    & \citet{Montet2015}         & 15.3353  & \phn57.2   & \phn0.7  & 2.4684        & \phn0.7     & 1.9  & 0.24\phn\phn\phn\phn  \\
201862715   & CP    & \citet{Burnham1882}        & 15.3350  & \phn99.7   & \phn0.4  & 1.4786        & \phn0.4     & 0.8  & 0.062\phn\phn\phn     \\
201890494   & EBC   & SDSS                       & 15.3379  & 256.6      & \phn1.3  & 2.7597        & \phn1.3     & 1.3  & 0.071\phn\phn\phn     \\
206061524   & PHOI  & This paper (Keck AO)       & \nodata  & 179.3      & \phn3.8  & 0.43\phn\phn  & 10\phd\phn  & 1.6  & 0.0035\phn\phn        \\ 
206135267   & EB    & This paper (Keck AO)       & \nodata  & 279.9      & \phn6.1  & 0.70\phn\phn  & 10\phd\phn  & 0.1  & 0.0088\phn\phn        \\ 
206152015   & EBC   & This paper (Keck AO)       & \nodata  & 291.4      & \phn2.2  & 0.25\phn\phn  & 10\phd\phn  & 1.1  & 0.0011\phn\phn       
\enddata
\tablecomments{The error on $\Delta m$ is 0.1~mag.}
\tablenotetext{a}{The tangential error.}
\tablenotetext{b}{For SOAR detections $\Delta m=\Delta m_I$.  For Keck detections, $\Delta m=\Delta m_{K_s}$.}
\tablenotetext{c}{This is the probability that projection effects could place an unbound background or foreground star at an angular separation less than or equal the measured separation.  See Section~\ref{subsec:associated} for more details.}
\label{tab:detect}
\end{deluxetable*}

\subsection{Non-detections}
\label{subsec:nondetect}

We discovered companion stars for only 12\% of our targets. However, non-detections are as important as detections in determining multiplicity rates. Due to distortions when measuring detection limits around binaries, we place detection limits only on the non-detections.  We estimated detection limits by the standard technique of calculating root-mean-square intensity fluctuations, $\sigma$, in annular zones of increasing radii and assumed that a companion with a central intensity of $5\sigma$ would be detectable.  For the SOAR non-detections, we also verified detection limits by simulating $\approx100$ companions near the expected $5\sigma$ detection curve for each star and attempting to recover them, typically validating the $5\sigma$ initial estimate for the detection curve, although it appeared to be a slightly conservative estimate.  Overall, the detection curves are more accurate at larger separations as the area of the annulus becomes larger.  The deeper, binned exposures gave better detections at large separations up to $3\arcsec$.

\begin{figure}
\includegraphics{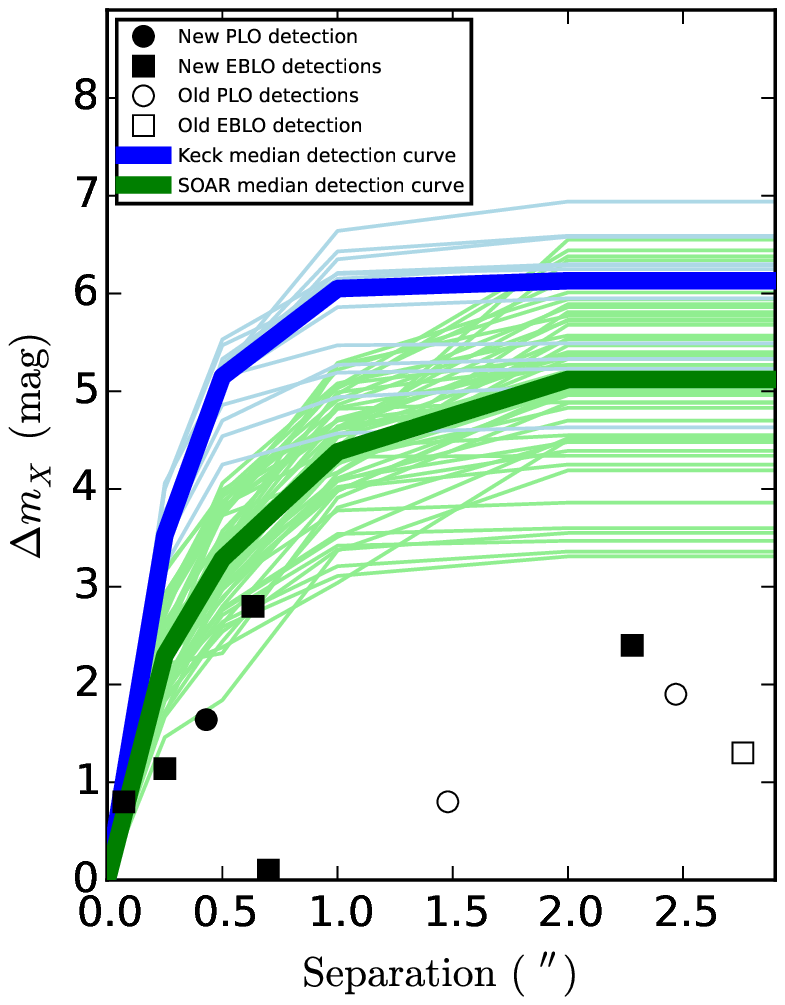}
\caption{Detection curve for every star with no detections using data from Table~\ref{tab:limits}.  Light blue lines are for Keck AO ($K_s$-band),  and light green lines are for SOAR speckle interferometry ($I$-band).  The median of all detection curves from each telescope is highlighted with a thicker, darker line (top thick line for Keck, bottom thick line for SOAR). Circles denote companions near planet-like objects (PLOs), and squares denote companions near EB-like objects (EBLOs). Empty symbols are previously known companions, while filled circles are newly discovered companions. Companion stars below the detection curve were likely to be detected, while those above were unlikely to be detected.}
\label{fig:limits}
\end{figure}

Table~\ref{tab:limits} and Figure~\ref{fig:limits} show the detection curves for all 66 stars with no detections.  Figure~\ref{fig:limits} also shows the median detection curve and the separations and $\Delta m$'s for all detections from both instruments, both previously known and newly discovered.  Five of the companions are at sub-arcsecond separations.

\begin{deluxetable*}{lcccccc}
\tablewidth{0pt}
\tablecaption{Detection Limits of the 52 Stars with No Detections.}
\tablehead{
\colhead{EPIC}                         & 
\colhead{Telescope}                    & 
\colhead{Min. Sep.\tablenotemark{a}}   & 
\colhead{$\Delta m(0.25\arcsec)$}      & 
\colhead{$\Delta m(0.50\arcsec)$}      & 
\colhead{$\Delta m(1.00\arcsec)$}      & 
\colhead{$\Delta m(2.00\arcsec)$}      \\ 
\colhead{ID}                           & 
\colhead{}                             &
\colhead{(mas)}                        & 
\colhead{(mag)}                        &
\colhead{(mag)}                        &
\colhead{(mag)}                        &
\colhead{(mag)}                        }
\startdata
201160662  & SOAR  & 0.14  & 2.70  & 4.06  & 5.08  & 5.57  \\ 
201207683  & SOAR  & 0.13  & 3.15  & 3.96  & 5.03  & 6.15  \\ 
201208431  & SOAR  & 0.12  & 2.36  & 3.44  & 4.43  & 4.55  \\ 
201246763  & SOAR  & 0.13  & 1.93  & 2.97  & 4.26  & 5.01  \\ 
201253025  & SOAR  & 0.18  & 2.02  & 2.89  & 4.08  & 4.70  \\ 
201257461  & SOAR  & 0.11  & 2.46  & 3.45  & 4.62  & 5.73  \\ 
201270464  & SOAR  & 0.08  & 2.15  & 2.38  & 3.04  & 4.50  \\ 
201295312  & SOAR  & 0.11  & 2.45  & 3.84  & 5.22  & 5.94  \\ 
201338508  & SOAR  & 0.15  & 2.23  & 3.15  & 4.01  & 4.25  \\ 
201367065  & SOAR  & 0.11  & 2.58  & 3.78  & 5.05  & 6.44  \\ 
201384232  & SOAR  & 0.14  & 2.24  & 2.87  & 4.36  & 5.33  \\ 
201393098  & SOAR  & 0.17  & 2.01  & 3.08  & 4.64  & 4.99  \\ 
201403446  & SOAR  & 0.11  & 2.77  & 3.76  & 5.29  & 6.01  \\ 
201407812  & SOAR  & 0.14  & 2.92  & 3.87  & 4.83  & 5.27  \\ 
201408204  & SOAR  & 0.13  & 2.84  & 3.93  & 4.61  & 4.99  \\ 
201445392  & SOAR  & 0.17  & 2.32  & 3.00  & 3.78  & 3.86  \\ 
201458798  & SOAR  & 0.14  & 2.87  & 3.99  & 4.70  & 4.97  \\ 
201465501  & SOAR  & 0.13  & 2.35  & 3.46  & 4.33  & 4.34  \\ 
201488365  & SOAR  & 0.11  & 1.46  & 1.84  & 3.37  & 4.19  \\ 
201505350  & SOAR  & 0.10  & 2.50  & 3.73  & 5.07  & 5.35  \\ 
201516974  & SOAR  & 0.13  & 2.64  & 3.51  & 4.97  & 6.38  \\ 
201546283  & SOAR  & 0.16  & 2.64  & 2.99  & 3.98  & 5.00  \\ 
201549860  & SOAR  & 0.12  & 2.24  & 3.57  & 4.82  & 4.89  \\ 
201555883  & SOAR  & 0.14  & 2.05  & 2.78  & 3.54  & 3.60  \\ 
201567796  & SOAR  & 0.11  & 2.21  & 3.83  & 4.43  & 6.34  \\ 
201569483  & SOAR  & 0.12  & 2.34  & 3.06  & 4.61  & 5.53  \\ 
201576812  & SOAR  & 0.11  & 2.21  & 2.32  & 3.91  & 4.88  \\ 
201577035  & SOAR  & 0.12  & 2.32  & 3.49  & 4.28  & 5.54  \\ 
201594823  & SOAR  & 0.14  & 2.84  & 3.80  & 4.70  & 5.40  \\ 
201596316  & SOAR  & 0.13  & 2.19  & 3.40  & 4.99  & 5.27  \\ 
201613023  & SOAR  & 0.18  & 1.90  & 2.80  & 4.16  & 5.47  \\ 
201617985  & SOAR  & 0.18  & 2.23  & 2.97  & 4.02  & 4.39  \\ 
201626686  & SOAR  & 0.18  & 1.88  & 3.37  & 4.86  & 5.72  \\ 
201629650  & SOAR  & 0.13  & 2.45  & 3.38  & 5.27  & 5.77  \\ 
201635569  & SOAR  & 0.19  & 2.05  & 2.64  & 3.21  & 3.36  \\ 
201648133  & SOAR  & 0.12  & 3.18  & 3.92  & 4.91  & 5.86  \\ 
201649426  & SOAR  & 0.16  & 2.27  & 3.36  & 4.72  & 5.35  \\ 
201665500  & SOAR  & 0.14  & 2.37  & 3.04  & 4.29  & 4.83  \\ 
201702477  & SOAR  & 0.18  & 1.92  & 2.71  & 3.42  & 3.47  \\ 
201705526  & SOAR  & 0.11  & 2.93  & 3.81  & 4.63  & 5.89  \\ 
201711881  & SOAR  & 0.11  & 2.27  & 3.32  & 4.55  & 5.38  \\ 
201725399  & SOAR  & 0.16  & 1.70  & 3.01  & 4.10  & 4.48  \\ 
201736247  & SOAR  & 0.10  & 1.72  & 2.59  & 3.11  & 3.31  \\ 
201754305  & SOAR  & 0.17  & 1.87  & 2.54  & 3.38  & 3.55  \\ 
201779067  & SOAR  & 0.11  & 2.59  & 3.25  & 4.40  & 6.55  \\ 
201855371  & SOAR  & 0.19  & 1.83  & 2.62  & 3.98  & 5.03  \\ 
201912552  & SOAR  & 0.10  & 2.62  & 3.53  & 4.25  & 5.81  \\ 
201920032  & SOAR  & 0.12  & 2.03  & 3.04  & 4.53  & 5.21  \\ 
201928968  & SOAR  & 0.08  & 2.91  & 3.73  & 4.17  & 5.68  \\ 
201929294  & SOAR  & 0.21  & 1.67  & 2.44  & 3.81  & 4.49  \\ 
203533312  & SOAR  & 0.13  & 2.01  & 2.76  & 3.51  & 4.52  \\ 
204129699  & SOAR  & 0.09  & 2.44  & 3.15  & 4.11  & 4.96  \\ 
205924614  & Keck  & 0.06  & 4.00  & 5.47  & 6.03  & 6.07  \\ 
205985357  & Keck  & 0.06  & 3.77  & 4.86  & 5.19  & 5.23  \\ 
206029314  & Keck  & 0.06  & 4.06  & 5.15  & 5.47  & 5.49  \\ 
206038483  & Keck  & 0.06  & 4.02  & 5.53  & 6.20  & 6.25  \\ 
206047297  & Keck  & 0.07  & 3.46  & 5.04  & 6.35  & 6.58  \\ 
206082454  & Keck  & 0.06  & 3.60  & 5.28  & 6.21  & 6.30  \\ 
206135075  & Keck  & 0.06  & 3.51  & 5.15  & 6.07  & 6.19  \\ 
206155547  & Keck  & 0.06  & 3.21  & 4.25  & 4.57  & 4.63  \\ 
206173295  & Keck  & 0.06  & 3.51  & 5.10  & 5.86  & 5.95  \\ 
206245553  & Keck  & 0.06  & 3.44  & 5.19  & 6.43  & 6.59  \\ 
206247743  & Keck  & 0.06  & 3.53  & 5.21  & 6.64  & 6.94  \\ 
206311743  & Keck  & 0.06  & 3.55  & 5.33  & 6.15  & 6.29  \\ 
206380678  & Keck  & 0.07  & 3.44  & 4.54  & 4.94  & 5.05  \\ 
206432863  & Keck  & 0.06  & 3.52  & 4.70  & 5.27  & 5.33  \enddata
\tablecomments{$\Delta m$ limits at $0.25\arcsec$, $0.50\arcsec$, $1.00\arcsec$, and $2.00\arcsec$. Beyond $2.00\arcsec$, the detection limits remain constant.  The SOAR observations were taken in the $I$ band, and the Keck observations were imaged in the $K_s$ band.}
\tablenotetext{a}{Minimum separation, approximately the distance at which a companion star with $\Delta m=0$ would be detectable.}
\label{tab:limits}
\end{deluxetable*}

\subsection{Physical association of the detected companions}
\label{subsec:associated}

Detected companions may be either physically bound to the primary star or may be a foreground or background star.  We tested the probability that any of our detections could be the result of a chance alignment with a non-physically associated star.  We used the \texttt{TRILEGAL} Galactic population model \citep{Girardi2005} to simulate a one square degree Galactic population of stars in the direction of each target with a detected companion and created nine simulated fields, one for each star with a companion.  We assumed that the distribution over this one square degree was uniform.  All of our detected companion stars from SOAR are brighter than $m_I=15.0$~mag, and all of our detected companion stars from Keck are brighter than $m_{K_s}=16.0$~mag, so we counted the number of brighter stars in each respective field (i.e., brighter than $m_I=15.0$~mag for SOAR fields and brighter than $m_{K_s}=16.0$~mag for Keck fields).  We then divided that number of stars by one square degree to get a surface density of stars and then multiplied by our FOV to determine the probability that any of these stars would be within our FOV.  For the nine stars, the probability of chance projection within $3\arcsec$ of the primary ranges between $0.07\%$ and $0.5\%$.  The probabilities are even lower when considering separations less than or equal to the measured companion separations rather than the entire $3\arcsec$ range (see Table~\ref{tab:detect}), strongly suggesting that all detected companions at these high Galactic latitudes are physically associated with their respective primaries.

\section{Discussion}
\label{sec:discussion}

High-resolution imaging is particularly important for exoplanets studies.  If a companion star is detected, it means that the signal from the planet is diluted and that the true planet radius is larger than initially measured.  The magnitude of this increase depends on the relative brightness of the two stars and knowledge of which star the planet orbits.  If the two stars are of near equal brightness, the true planet radius will be about half that which was measured.  For binaries with a large $\Delta m$, the true planet radius will either be nearly the same as the measured value (if the planet orbits the primary star), or the true planet radius will be greatly increased (if the planet orbits the secondary star).  A good example of correcting for the dilution caused by a companion star is shown in \citet{Dressing2014}.  The average planetary radius from transit surveys may be underestimated by a factor of 1.5, though this can be reduced to 1.2 with radial velocity and high-resolution data \citep{Ciardi2015}. 

These data are also useful for the statistical validation of planet candidates.  Both detections and non-detections with contrast curves can provide sufficient constraints to rule out enough parameter space from astrophysical false positives to statistically validate the planet candidate as a true planet.  This has been done for many planets with the \texttt{BLENDER} code \citep[e.g.,][]{Torres2011}.

Recently, studies have also attempted to determine the relationship between stellar multiplicity and exoplanets.  The multiplicity rate of known exoplanet hosts compared to stars not known to host planets can inform our knowledge of planet formation\footnote{One must be careful not to say ``known non-exoplanet host stars'', as it is currently impossible to prove that any one star does not host any planet.  Rather, one usually compares known exoplanet host stars to stars known not to host stars above some detectable threshold or to field stars, some of which will host undiscovered exoplanets.}.  If exoplanets are more frequently found in multiple star systems, one can assume that multiplicity enhances planet formation. If exoplanets are found to be less common in multiple star systems, one can conversely assume that multiplicity suppresses planet formation.  Studies differ on whether the multiplicity rate of known exoplanet host stars is consistent with the multiplicity rate of stars without known exoplanets \citep{Bonavita2007, Raghavan2010, Lodieu2014} or whether the multiplicity rate of known exoplanet host stars is lower \citep{Mugrauer2009, Roell2012, Wang2014b}.  The existence of companion stars may also influence the architecture of the planetary system \citep[e.g.,][]{Desidera2007,Quintana2007,Roell2012}, although some studies have put constraints on their potential influence, such as no correlation existing between misaligned or eccentric hot Jupiters and the incidence of directly imaged stellar companions \citep[e.g.,][]{Ngo2015}.

\section{Conclusions} 
\label{sec:conclusion}

We found nine companion stars within $3\arcsec$ of three candidate transiting exoplanet host stars and six EB candidates.  All nine companion stars are likely to be physically associated with the target star.  Six of the nine detected companions are new discoveries. Five of these six companions are associated with likely EBs. 

Without knowledge of the physical binary separations, it is difficult to determine whether or not there are any potentially significant deviations between the binary statistics in any sub-sample of our target stars and the binary statistics of the population of field stars.  However, it is worth noting that many of the short-period EBs and EB candidates ($P<3$~days) were found to have companions, supporting the conclusions in \citet{Tokovinin2006}that all short-period ($P<3$~days) EBs have wider companions.  These observations contribute to the growing data set describing the multiplicity of our galactic neighborhood. This will soon help shed light on the influence that stellar multiplicity might have on planet formation.   

\vspace{0.25in}\noindent\textit{Acknowledgements} 

J.R.S. and T.S.B. acknowledge support from NASA ADAP 14-0245.  D.A.F. acknowledges funding support for Planet Hunters from Yale University and acknowledges support from NASA ADAP12-0172. T.S.B. acknowledges funding support from 14-K2GO1\_2-0075, 14-K2GO2\_2-0075, and 15-K2GO3\_2-0063.  K.S. gratefully acknowledges support from Swiss National Science Foundation Grant PP00P2\_138979/1. The Zooniverse is supported by The Leverhulme Trust and by the Alfred P. Sloan foundation. PH is supported in part by NASA JPL's PlanetQuest program. The data presented in this paper are the result of the efforts of the PH volunteers, without whom this work would not have been possible. Their contributions are individually acknowledged at \url{http://www.planethunters.org/authors}. The authors thank the PH volunteers who participated in identifying and analyzing the planet and EB candidates presented in this paper.  The authors also thank Andrew Vanderburg and the Harvard-Smithsonian Center for Astrophysics for making available the reduced light curves for K2 C1, C2, and C3.

Some of the research presented in this paper is based on observations obtained at the Southern Astrophysical Research (SOAR) telescope, which is a joint project of the Minist$\rm{\acute{e}}$rio da Ci$\rm{\hat{e}}$ncia, Tecnologia, e Inova\c{c}$\rm{\tilde{a}}$o (MCTI) da Rep$\rm{\acute{u}}$blica Federativa do Brasil, the U.S. National Optical Astronomy Observatory (NOAO), the University of North Carolina at Chapel Hill (UNC), and Michigan State University (MSU).  Some of the data presented herein were obtained at the W. M. Keck Observatory, which is operated as a scientific partnership among the California Institute of Technology, the University of California and the National Aeronautics and Space Administration.  The Observatory was made possible by the generous financial support of the W. M. Keck Foundation.

This paper includes data collected by the \textit{Kepler} spacecraft, and we gratefully acknowledge the entire \textit{Kepler} mission team's efforts in obtaining and providing the light curves used in this analysis. Funding for the \textit{Kepler} mission is provided by the NASA Science Mission directorate. Support for MAST for non-HST data is provided by the NASA Office of Space Science via grant NNX13AC07G and by other grants and contracts.  This research has made use of NASA's Astrophysics Data System Bibliographic Services, the Washington Double Star Catalog maintained at the U.S. Naval Observatory, and the APASS database, located at the AAVSO website. Funding for APASS has been provided by the Robert Martin Ayers Sciences Fund. This publication makes use of data products from the Two Micron All Sky Survey, which is a joint project of the University of Massachusetts and the Infrared Processing and Analysis Center/California Institute of Technology, funded by the National Aeronautics and Space Administration and the National Science Foundation.  Funding for the Sloan Digital Sky Survey IV has been provided by the Alfred P. Sloan Foundation, the U.S. Department of Energy Office of Science, and the Participating Institutions. SDSS- IV acknowledges support and resources from the Center for High-Performance Computing at the University of Utah. The SDSS website is \url{www.sdss.org}.   This publication makes use of data products from the \textit{Wide-field Infrared Survey Explorer}, which is a joint project of the University of California, Los Angeles, and the Jet Propulsion Laboratory/California Institute of Technology, funded by the National Aeronautics and Space Administration.

\bibliographystyle{apj}
\bibliography{bib}

\appendix

Determining the true planet occurrence rate requires knowledge of the sample's selection effects.  Therefore, we provide here a table of the selection biases for each star as determined by the GO proposals which requested that the star be observed so that any potential future analysis of the planet occurrence rate using these stars can attempt to account for these biases.  

\section{Selection biases}

\clearpage
{\LongTables
\begin{deluxetable*}{lccl}
\tablewidth{0pt}
\tabletypesize{\footnotesize}
\tablecaption{Selections biases in target selection.}
\tablehead{
\colhead{EPIC}                                              & 
\colhead{Status}                                            & 
\colhead{Detection\tablenotemark{a}}                        &
\colhead{General Selection Biases in Order of GO Proposals\tablenotemark{b}} }
\startdata
201160662  & EBC\phm{C}  &                    & Late-FGK dwarfs  \\
201207683  & EBC\phm{C}  &                    & Red giants, but with overlap from KM dwarfs  \\ 
201208431  & VP\phm{CC}  &                    & Red giants, but with overlap from KM dwarfs  \\ 
201246763  & EBC\phm{C}  &                    & Late-FGK dwarfs  \\ 
201253025  & EB\phm{CC}  &                    & Late-FGK dwarfs | Known EBs   \\ 
201257461  & FP\phm{CC}  &                    & Red giants, but with overlap from KM dwarfs  \\ 
201270464  & EBC\phm{C}  &                    & Metallic-line A stars | A0-F5 with a peculiar chemical composition, with pulsations, or \\
&&& in multiple star systems | A to early-F stars  \\ 
201295312  & PC\phm{CC}  &                    & Late-FGK dwarfs  \\ 
201324549  & EBC\phm{C}  & $\checkmark$       & Late-FGK dwarfs  \\ 
201338508  & VP\phm{CC}  &                    & Late-FGK dwarfs | Red giants, but with overlap from KM dwarfs  \\ 
201367065  & VP\phm{CC}  &                    & M-dwarfs | M-dwarfs | M-dwarfs (M0-M5) | M-dwarfs | Red giants, but with \\ 
&&&overlap from KM dwarfs | M-dwarfs | M-dwarfs (M0-M5)  \\ 
201384232  & VP\phm{CC}  &                    & Late-FGK dwarfs  \\ 
201393098  & VP\phm{CC}  &                    & Late-FGK dwarfs  \\ 
201403446  & PC\phm{CC}  &                    & Late-FGK dwarfs  \\ 
201407812  & EBC\phm{C}  &                    & Late-FGK dwarfs  \\ 
201408204  & EB\phm{CC}  &                    & Late-FGK dwarfs | Known EBs  \\ 
201445392  & PC\phm{CC}  &                    & Red giants, but with overlap from KM dwarfs  \\ 
201458798  & EBC\phm{C}  &                    & Late-FGK dwarfs  \\ 
201465501  & VP\phm{CC}  &                    & M-dwarfs | M-dwarfs | M-dwarfs, emphasizing M4 and later | Red giants, but with \\
&&& overlap from KM dwarfs | M-dwarfs, with the lower priority targets containing some \\
&&&likely M5-M8 dwarfs  \\ 
201488365  & EB\phm{CC}  &                    & Known EBs (eclipsing Algols, EBs of the beta Lyr type, and EBs of the W Uma type) \\
&&&| Known EBs | Late-FGK dwarfs | Known EBs | F-dwarfs  \\ 
201505350  & CP\phm{CC}  &                    & Late-FGK dwarfs  \\ 
201516974  & PHOI        &                    & Late-FGK dwarfs | Red giants, but with overlap from KM dwarfs  \\ 
201546283  & PC\phm{CC}  &                    & Late-FGK dwarfs  \\ 
201549860  & PC\phm{CC}  &                    & Red giants, but with overlap from KM dwarfs  \\ 
201555883  & FP\phm{CC}  &                    & Red giants, but with overlap from KM dwarfs  \\ 
201567796  & EBC\phm{C}  &                    & Late-FGK dwarfs  \\ 
201569483  & FP\phm{CC}  &                    & Late-FGK dwarfs | Red giants, but with overlap from KM dwarfs  \\ 
201576812  & EB\phm{CC}  &                    & Late-FGK dwarfs | GKM dwarfs | Known EBs   \\ 
201577035  & VP\phm{CC}  &                    & Late-FGK dwarfs  \\ 
201594823  & EB\phm{CC}  &                    & Late-FGK dwarfs | Known EBs   \\ 
201596316  & VP\phm{CC}  &                    & Late-FGK dwarfs | Red giants, but with overlap from KM dwarfs  \\ 
201613023  & VP\phm{CC}  &                    & Late-FGK dwarfs  \\ 
201617985  & PC\phm{CC}  &                    & M-dwarfs (M0-M6) with no 2MASS object within $10\arcsec$ | Red giants, but with overlap from \\
&&& KM dwarfs  \\ 
201626686  & EBC\phm{C}  &                    & A to early-F stars | Late-FGK dwarfs  \\ 
201629650  & VP\phm{CC}  &                    & Late-FGK dwarfs  \\ 
201635569  & VP\phm{CC}  &                    & M-dwarfs, emphasizing M4 and later | Red giants, but with overlap from KM dwarfs  \\ 
201648133  & EBC\phm{C}  &                    & Late-FGK dwarfs  \\ 
201649426  & FP\phm{CC}  &                    & Red giants, but with overlap from KM dwarfs  \\ 
201665500  & EB\phm{CC}  &                    & Late-FGK dwarfs | Known EBs   \\ 
201702477  & PC\phm{CC}  &                    & Red giants, but with overlap from KM dwarfs  \\ 
201704541  & EB\phm{CC}  & $\checkmark$       & Known EBs  \\ 
201705526  & EBC\phm{C}  &                    & Late-FGK dwarfs | F-dwarfs  \\ 
201711881  & EB\phm{CC}  &                    & Cepheids | Late-FGK dwarfs  \\ 
201725399  & EBC\phm{C}  &                    & Known EBs  \\ 
201736247  & VP\phm{CC}  &                    & Red giants, but with overlap from KM dwarfs  \\ 
201754305  & VP\phm{CC}  &                    & Red giants, but with overlap from KM dwarfs  \\ 
201779067  & FP\phm{CC}  &                    & Late-FGK dwarfs | GKM dwarfs  \\ 
201826968  & EBC\phm{C}  & $\checkmark$       & Late-FGK dwarfs  \\ 
201828749  & PC\phm{CC}  & $\checkmark$       & Late-FGK dwarfs  \\ 
201855371  & VP\phm{CC}  &                    & Red giants, but with overlap from KM dwarfs  \\ 
201862715  & CP\phm{CC}  & $\checkmark$       & Binaries from WDS with separation $<1.5\arcsec$ | Solar-like planet-hosting stars | \\ 
&&&WASP-85 (late-FGK dwarfs)\tablenotemark{c} | Late-FGK dwarfs | Red giants, but with overlap from \\
&&& KM dwarfs  \\ 
201890494  & EBC\phm{C}  & $\checkmark$       & Late-FGK dwarfs  \\ 
201912552  & VP\phm{CC}  &                    & M-dwarfs | M-dwarfs | M-dwarfs (M0-M5) | M-dwarfs (M0-M4) | M-dwarfs | \\
&&&M-dwarfs | Red giants, but with overlap from KM dwarfs | M-dwarfs (M0-M5)  \\ 
201920032  & PHOI        &                    & A0-F5 with a peculiar chemical composition, with pulsations, or in multiple star systems | \\
&&& Late-FGK dwarfs | F-dwarfs  \\ 
201928968  & EBC\phm{C}  &                    & Proper motion selected wide binaries $>5\arcsec$ and $<120\arcsec$  \\ 
201929294  & FP\phm{CC}  &                    & Red giants, but with overlap from KM dwarfs  \\ 
203533312  & EB\phm{CC}  &                    & Red giants, but with overlap from KM dwarfs | Late-FGK dwarfs | FGK dwarfs  \\ 
204129699  & EBC\phm{C}  &                    & Late-FGK dwarfs | FGK dwarfs \\ 
205924614  & PHOI        &                    & Red giants, but with overlap from KM dwarfs \\ 
205985357  & EBC\phm{C}  &                    & Red giants, but with overlap from KM dwarfs \\ 
206029314  & EBC\phm{C}  &                    & Late-FGK dwarfs \\ 
206038483  & PHOI        &                    & FGK dwarfs | Late-GFK dwarfs |  A0-F5 stars that might be stars with a peculiar \\ 
&&&chemical composition, pulsating stars, or multiple star systems \\ 
206047297  & PHOI        &                    & Red giants, but with overlap from KM dwarfs \\ 
206061524  & PHOI        & $\checkmark$       & Late-FGK dwarfs | Red giants, but with overlap from KM dwarfs | M-dwarfs \\ 
206082454  & PHOI        &                    & FGK dwarfs | Late-FGK dwarfs \\ 
206135075  & PHOI        &                    & FGK dwarfs | Late-FGK dwarfs |  A0-F5 stars that might be stars with a peculiar \\
&&&chemical composition, pulsating stars, or multiple star systems \\ 
206135267  & EB\phm{CC}  & $\checkmark$       & FGK dwarfs | Late M-dwarf EBs | GKM dwarfs | Late-FGK dwarfs | Red giants, \\
&&&but with overlap from KM dwarfs |  A0-F5 stars that might be stars with a peculiar \\ 
&&&chemical composition, pulsating stars, or multiple star systems \\ 
206152015  & PHOI        & $\checkmark$       & Late-FGK dwarfs \\ 
206155547  & PHOI        &                    & GKM dwarfs \\ 
206173295  & PHOI        &                    & Late-FGK dwarfs \\ 
206245553  & PHOI        &                    & FGK dwarfs | Late-FGK dwarfs \\ 
206247743  & PHOI        &                    & Red giants, but with overlap from KM dwarfs \\ 
206311743  & EBC\phm{C}  &                    & FGK dwarfs | Late-FGK dwarfs | Red giants, but with overlap from KM dwarfs \\ 
206380678  & PHOI        &                    & Late-FGK dwarfs | Red giants, but with overlap from KM dwarfs \\ 
206432863  & PHOI        &                    & Late-FGK dwarfs
\enddata
\tablecomments{CP is defined as confirmed planet, VP as validated planet, PC as planet candidate, PHOI as Planet Hunters Object of Interest, FP as fall positive, EB as a previously known eclipsing binary, and EBC as an EB candidate.  All VPs, PCs, and FPs, and one CP (EPIC 201505350) are from \citet{Montet2015}, while all other stars are from Planet Hunters.}
\tablenotetext{a}{Detected in this paper.}
\tablenotetext{b}{Selection biases between different GO proposals for the same star are separated by ``|'' in the same order as listed on the K2 website (Campaign 1: \url{http://keplerscience.arc.nasa.gov/K2/GuestInvestigationsC01.shtml} and Campaign 2: \url{http://keplerscience.arc.nasa.gov/K2/GuestInvestigationsC02.shtml}).}
\tablenotetext{c}{EPIC 201862715 was originally selected for by the WASP team based on its classification as a late-FGK dwarf.  Its binarity was not taken into account for its selection (D. Brown 2015, private communication).}
\label{tab:bias}
\end{deluxetable*}
}

\end{document}